\documentclass[%
 aip,
 apl,
 amsmath,amssymb,
reprint,%
]{revtex4-1}

\usepackage{graphicx}
\usepackage{dcolumn}
\usepackage{bm}

\usepackage[separate-uncertainty=true]{siunitx}
\usepackage[version=4]{mhchem}
\usepackage{float}
\usepackage{mathtools}
\usepackage{hyperref}
\DeclareSIUnit\samples{\text{S}}

\usepackage{graphicx}
\usepackage{dcolumn}
\usepackage{bm}

\usepackage[utf8]{inputenc}
\usepackage[T1]{fontenc}
\usepackage{mathptmx}
\usepackage{etoolbox}

\makeatletter
\def\@email#1#2{%
 \endgroup
 \patchcmd{\titleblock@produce}
  {\frontmatter@RRAPformat}
  {\frontmatter@RRAPformat{\produce@RRAP{*#1\href{mailto:#2}{#2}}}\frontmatter@RRAPformat}
  {}{}
}%
\makeatother

\begin{document}
    \preprint{AIP/123-QED}
    \title{Markov chain Monte Carlo Detector Tomography applied to a NbTiN nanobridge}
    
    \author{F.B. Baalbergen}
    \email{baalbergen@physics.leidenuniv.nl}
    \affiliation{Huygens-Kamerlingh Onnes Laboratory, Leiden University, P.O. Box 9504, 2300RA Leiden, The Netherlands}
    \author{I.E. Zadeh}
    \affiliation{Department of Imaging Physics (ImPhys), Faculty of Applied Sciences, Delft University of Technology, 2628CJ Delft, The Netherlands}
    
    \author{M.P. van Exter}
    \affiliation{Huygens-Kamerlingh Onnes Laboratory, Leiden University, P.O. Box 9504, 2300RA Leiden, The Netherlands}
    
    \author{M.J.A. de Dood}
     \email{dood@physics.leidenuniv.nl}
    \affiliation{Huygens-Kamerlingh Onnes Laboratory, Leiden University, P.O. Box 9504, 2300RA Leiden, The Netherlands}
    
    \date{\today}
    
    \begin{abstract}
        We demonstrate the use of a flexible and highly accurate Markov chain Monte Carlo Quantum Detector Tomography method as a minimization algorithm to best describe the response of an efficient \qty{120}{\nm} wide \ce{NbTiN} superconducting nanobridge single photon detector.
        Separation of the internal quantum efficiency and external quantum efficiency is possible due to the difference in saturation behavior of an ideal 1-photon threshold detector as compared to a detector with non-unity 1-photon internal quantum efficiency.
        From a statistical analysis of our measurements (at $T=\qty{4.23(1)}{\kelvin}$, $I=\qty{29.4(1)}{\micro\ampere}$, $I/I_\text{switch}=\num{0.90(1)}$) we find an external quantum efficiency $\eta=\num{1.60(5)e-6}$, a 1-photon internal quantum efficiency $p_1=\num{0.568(8)}$ and unity multi-photon (2 or more) internal quantum efficiency.
    \end{abstract}
    
    \maketitle

    Superconducting nanowire single photon detectors (SNSPDs) are one of the most successful detector technologies for quantum optics, quantum state preparation and quantum information applications.
    Unlike other superconducting detector technologies, such as transition edge sensors (TES), kinetic inductance detectors (KIDs) and superconducting tunnel junctions (STJs), SNSPDS can operate above \qty{2}{\kelvin} enabling the use of compact cryocoolers~\cite{Verma2014, Miki2009}, and operate at significant bias currents close to the device switching current.
    The operation principle enables devices that display a subset of fast recovery times, low timing jitter, high detection efficiencies~\cite{Chang2021,Reddy2020,Hu2020}, low dark count rates\cite{Subashchandran2013, Shibata2015} as well as intrinsic photon number resolution~\cite{Zhu2020, Los2024}.
    These properties make SNSPDs an attractive detector technology for future quantum applications.

    Detector characterization of optical detectors is challenging due to large uncertainties in optical power measurements.
    Furthermore, it is hard to disentangle loss in the system from intrinsic efficiencies.
    In this work we use quantum detector tomography to characterize both the internal and external quantum efficiency of a nanobridge detector.
    The purpose of quantum detector tomography is to find the detection probabilities $p_i$ in the photon number basis (Fock states) in an agnostic way.
    To achieve this we use coherent states as a tomographically complete set. 
    We emphasize the importance of the quality (low noise, unbiased) of the raw data measured over a sufficiently large dynamic range for correct tomographic estimation of detector parameters. 
    This data can be analyzed using statistical methods that assess goodness of fit to distinguish between detectors with 100\% internal quantum efficiency, i.e.\ ideal one-photon threshold detectors where every absorbed photon generates an electronic click, and more realistic detectors that have a less than ideal internal quantum efficiency.
    Understanding small differences in internal quantum efficiency is key when characterizing detectors that operate at the edge of what is possible with SNSPDs.
    In particular we mention high-speed operation, (mid)-infrared wavelengths and near-unity detection efficiency for applications in photonic quantum computing. 
    
    Coherent states are readily available as the output from a (pulsed) laser source.
    For a single photon threshold detector illuminated by a coherent state with $\left<n\right>$ photons per pulse, the probability for an $i$\nobreakdash-photon detection event is given by~\cite{Lundeen2009,Renema2012}
    \begin{align}
        P_{i,\text{click}}&=p_i\exp(-\mu_\text{eff})\frac{\mu_\text{eff}^i}{i!}.
    \end{align}
    Here we have introduced the effective photon number $\mu_\text{eff}=\eta \left<n\right>$, where $\eta$ is the overall external probability to absorb a photon in the active area.
    The internal quantum efficiencies $p_i$ are the probabilities that the absorption of $i$ photons generate an electronic click~\cite{Bienfang2023}.
    By summing the probabilities for all photon counts and by using the probability of the detector not clicking, the total probability can be described as
    \begin{align}
        P_\text{click}&=1-P_\text{no click}=1-\mathrm{e}^{-\mu_\text{eff}}\sum_{i=0}^{n_\text{max}}\left(1-p_i\right)\frac{{\mu_\text{eff}}^i}{i!}.\label{eq:tomography}
    \end{align}
    The summation is truncated at a photon number threshold $i = n_\text{max}$.
    Above this threshold the detection probabilities $p_i$ are equal to 1.
    
    We note that an ideal single photon threshold detector with no dark counts corresponds to $p_0=0$ and $p_i=1$ for all $i\ge1$.
    In this case characterizing the detection with a single quantum efficiency $\eta$ suffices.
    Ultimately the goal of SNSPDs is to achieve near-100\% detection efficiency over a broad wavelength range extending towards the mid-infrared.
    A tool that can quantify the internal quantum efficiency is essential to gain an understanding of the physics that prevents ideal single-photon threshold detection. 
    
\begin{figure}
    \centering
    \includegraphics[width=85mm]{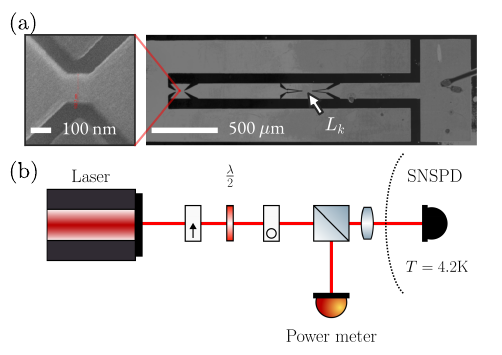}
    \caption{
        \textbf{(a)} SEM image of the nanobridge detector (left) and optical image showing the complete device with the nanobridge and inductor ($L_k$, right). 
        \textbf{(b)} Schematic of the setup showing laser, power meter and detector.
        The pulsed laser is attenuated using a computer controlled waveplate and crossed polarizers.
    }\label{fig:methods}
\end{figure}

    Our \ce{NbTiN} superconducting nanobridge single photon detectors are fabricated using standard e\nobreakdash-beam lithography and etching techniques~\cite{Zichi2019} on a \qty{230}{\nm} thick layer of \ce{SiO2} on a \ce{Si (100)} wafer.
    The \ce{NbTiN} superconductor film is \qty{13}{\nm} thick and is capped using a \qty{12}{\nm} thick \ce{Si3N4} layer.
    The nanobridge detector is defined by creating a constriction  of \qtyproduct{120 x 120}{\nano\meter} in a \qty{500}{\nm} wide wire.
    The \qty{500}{\nm} wide wire is extended into a meandering wire to create  $\sim\qty{ 700}{\nano\henry}$ on-chip inductance.
    This inductance is connected in series with the nanobridge to slow down the response of the detector and prevent latching~\cite{Annunziata20102}.
    The detectors are mounted in a closed cycle cryostat with optical access and are cooled to \qty{4.23(1)}{\kelvin}.
    An optical microscope image of the device and SEM image of the nanobridge, is shown in Fig.~\ref{fig:methods}a.

    The detectors are biased using a quasi-constant voltage bias~\cite{Liu2012,baalbergen2024}.
    A voltage source (Yokogawa GS200) is connected in series to a \qty{10}{\kilo\ohm} resistor. 
    Parallel to the DC input of the bias tee (minicircuits ZFBT\nobreakdash-6\nobreakdash-GW+) a \qty{50}{\ohm} resistor to ground is connected.
    Together with the series inductance of the meander this resistor prevents latching of the detector.
    The AC input of the bias tee is connected to two room temperature, AC-coupled amplifiers (minicircuits ZFL\nobreakdash-1000LN+).

    The detector is illuminated using a pulsed laser (Id Quantique ID-3000) with a $f_\text{rep}=\qty{5}{\mega\hertz}$ repetition rate.
    The laser produces $\lambda=\qty{850}{\nm}$ pulses with an approximately \qty{50}{\pico\second} pulse width.
    The average optical power is attenuated between \qty{5}{\nano\watt} and \qty{7}{\micro\watt} using a rotatable $\lambda/2$\nobreakdash-plate placed between two crossed Glan\nobreakdash-Thompson polarizing prisms (B. Halle).
    A beam splitter sends approximately half of the optical power to a power meter that continuously probes the average laser power during the tomography experiment.
    A second $\lambda/2$\nobreakdash-plate sets the linear polarization of the incident light to maximize the detection count rate, i.e.\ $E$-field parallel to the long axis of the \ce{NbTiN} wire\cite{Renema2015}.
    A schematic overview of the optics is shown in Fig.~\ref{fig:methods}b.

    In order to find the experimental click probability $P_\text{click}$, the output of the detector is amplified and recorded using a digitizer (Teledyne ADQ7, \qty{5}{\giga\samples\per\second}, \qty{3}{\giga\hertz} analog bandwidth).
    The digitizer is used to measure the time delay $\Delta t$ between the rising edge of the laser synchronization signal and the maximum derivative of the SNSPD detector pulse.
    We define a detection pulse as a light pulse when the time delay falls within a \qty{2}{\nano\second} window after the light reaches the detector.
    This window size is chosen as the complete width of the histogram of all measured time delays to ensure that all light pulses are incorporated.
    All pulses that fall outside this window are seen as dark counts and are therefore not included in the calculation of $P_\text{click}$.

    The nanobridge detector is characterized at a fixed current of \qty{29.4(1)}{\micro\ampere}.
    At the used temperature of \qty{4.23(1)}{\kelvin}, the switching current of the device is \qty{32.8(1)}{\micro\ampere}.
    For each power setting, the data for approximately \num{3e5} laser pulses was measured by the digitizer.
    To avoid issues with long term stability and laser fluctuation we vary the power in our experiment rapidly; measurements at different powers are only \qty{2}{\second} apart~\cite{Wang2015}.

    To perform detector tomography we employ Markov chain Monte Carlo (MCMC) likelihood maximization to estimate the best parameters $\{\eta,p_i\}$ in equation~\eqref{eq:tomography} for the measured data~\cite{Foreman-Mackey2013, Foreman2019}.
    We improve on previously published results\cite{Renema2012,Wang2015} in the following ways: 
    The quality of the data has been improved by monitoring the laser power with a calibrated power monitor to cancel laser intensity fluctuations. 
    The click probability is measured at lower repetition rates, and all detection events are timed relative to the laser pulses.
    This timing information allows to filter out more than $98\%$ of dark events.
    The MCMC algorithm makes better estimates of the (non-Gaussian) probability distributions and covariance of the various parameters when finding the best parameters in the tomography model.
    This makes it possible to obtain a good fit of the tomography model to the data at  both the very low and very high photon fluxes.
    In the low photon flux regime dark counts would be dominant over light counts, whereas in the high photon flux regime dark counts would lead to the possibility of measuring the non physical $P_\text{click}>1$.
    The improvements in the method of effective detector tomography allow to explore detectors with high internal detection efficiencies, i.e.\ with $p_1 \sim 1$.

    At each power input setting $j$ with average photon number $\left<n\right>_j\pm\sigma_{\left<n\right>_j}$ we measure the click probability $P_j\pm\sigma_{P_j}$.
    The errors on the photon numbers of the input state are determined by the measurement error as specified by the manufacturer ($3\%$).
    Since the error in the measured power (3\%) is multiple times the error caused by the non-linearity of the power meter (0.5\%), we only consider the measurement errors in the log likelihood.
    The MCMC method uses the likelihood of the model, computed by assuming Gaussian distributed, uncorrelated errors on the measurements.
    By assuming that the tomography model can locally be approximated to be linear, it is possible to derive an analytic expression for the log likelihood function.
    The log likelihood is then defined as a summation of all normalized orthogonal distances between the tomography model and the measurement data.
    Using this method, the log likelihood function is defined as
    \begin{align}
        \ell=
        -\frac{1}{2}
        \sum_j
        \Bigg(
            &{\left(\frac{\sigma_{P_j}}{P_\text{click}(\left<n\right>_j) - P_j}\right)}^2
            +\notag\\
            &{\left(\frac{\sigma_{\left<n\right>_j}}{P_\text{click}^{-1}(P_j) - \left<n\right>_j}\right)}^2
        \Bigg)^{-1}=-\frac{1}{2}\chi^2\label{eq:loglike}.
    \end{align}
    Where $P_\text{click}(n)$ and $P_\text{click}^{-1}(P)$ are the tomography model from equation~\ref{eq:tomography} and the inverse of the tomography model.
    This summation of distances is defined as $\chi^2$, the chi square of the fit of the model. 

\begin{figure}
    \centering
    \includegraphics[width=85mm]{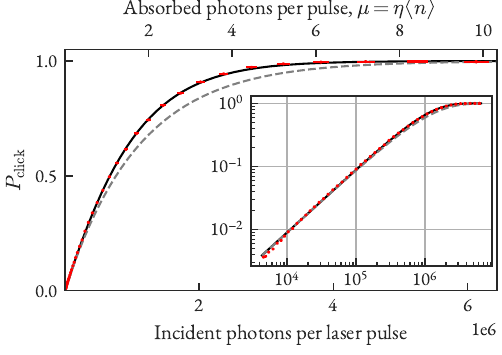}
    \caption{
        Measured click probability vs.\ average incident photons.
        The fit (solid line) to the data (red symbols), giving $\eta=\num{1.60(5)e-6}$ and $p_1=\num{0.568(8)}$, is compared to the click probability for a corresponding ideal single-photon threshold detector (dashed line, $\eta=\num{9.0(3)e-7}$, $p_{\geq1}=1$).
        When plotted on a linear axis, a clear difference between the two models is visible.
        Inset is the same data on log-log axes, showing the single-photon response and saturation to $P_\text{click}=1$. 
    }\label{fig:linecut}
\end{figure}

    Figure~\ref{fig:linecut} shows the measured click probability as a function of laser power (red points).
    The data is shown on a linear scale (main figure) and logarithmic scale (inset).
    The solid lines indicate the best fit of the model with $n_\text{max}=1$.
    This best fit to the data contains both external quantum efficiency $\eta$ and the internal quantum efficiency $p_1$ as fit parameters.
    The dashed lines show the expected detection probability for an ideal single photon threshold detector with $p_1=1.0$.
    The value for the product $\eta p_1$ is determined from the slope of the data in the low power regime (see inset).
    On a linear scale a clear difference between the tomography model and ideal single photon model is visible.
    
\begin{figure}
    \centering
    \includegraphics[width=85mm]{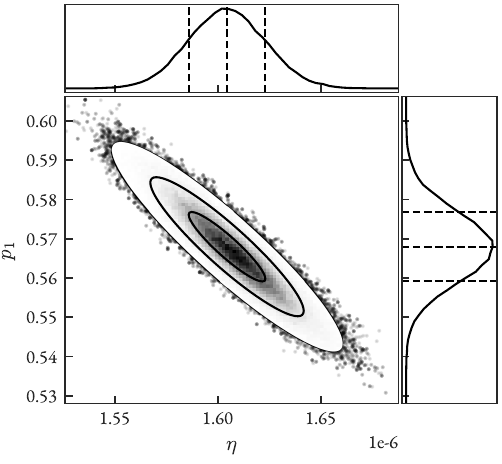}
    \caption{Analysis of the MCMC method showing the covariance between the parameters $\eta$ and $p_1$.
            The square figure shows an inverse correlation between both parameters.
            The plots above (besides) the square plot, show the distribution for $\eta$ ($p_1$).
            }\label{fig:correlation}
\end{figure}

    The MCMC method gives a direct estimate of the correlation in the fit parameters.
    These correlations for a 2 parameter model ($\eta$ and $p_1$) are shown in figure~\ref{fig:correlation}.
    As expected the parameters $\eta$ and $p_1$ are anti-correlated.
    The fit parameters are approximately distributed according to a multivariate normal distribution.
    For our detector we find an external quantum efficiency $\eta=\num{1.60(5)e-6}$ and internal quantum efficiency of $p_1=\num{0.568(8)}$.
    The error bar on $p_1$ is obtained from the variation in the log likelihood function.
    For the error bar on $\eta$, the statistical error as found from the variation in the log likelihood function is added to the systematic error of the power meter (3\%) with standard error propagation.

    While we corrected our data for dark counts, we have not corrected for the dead-time due to dark counts or after-pulsing effects~\cite{Raupach2023}.
    By using an $1/e$-time of \qty{14(2)}{\nano\second} and a dark count rate of \qty{1000}{\per\second} we can estimate that approximately $75$ times per second the detector is not active during the light pulse. 
    For our repetition rate, this gives an correction of $\sim \num{2e-4}$ of the measured counting probabilities.

    In the tomography model from equation~\ref{eq:tomography} there is a choice in the number of free parameters $p_i$.
    For our detector we found that using a single internal quantum efficiency leads to the best trade off between the amount of parameters and the goodness of fit.
    The goodness of fit is determined by using the reduced $\chi^2_r$ where $\chi^2_r=\chi^2/\text{DOF}$ where $\text{DOF}$ is the amount of degrees of freedom of the fit.
    When comparing the best models with two and three parameters, the $\chi^2_r\approx4.8$ for the two parameter model ($\eta$ and $p_1$) is slightly lower then $\chi^2_r\approx5.0$ for the three parameter model ($\eta$, $p_1$ and $p_2$) consistent with model selection based on an Akaike Information Criterion\footnote{It is also possible to use the Aikaike Information Criterion (AIC) for the selection of the best amount of model parameters. The AIC is defined as $AIC=2k-2\ell=2k+\chi^2$ with $k$ the number of estimated parameters~\cite{Akaike1974}. For our models we find for a one parameter model $AIC=955$, for a two parameter model $AIC=236$, and finally for a three parameter model $AIC=238$ favoring the model with 2 fit parameters.}.
    Moreover, in the best three parameter model, it is found that $p_2=0.98^{+0.02}_{-0.03}$ with $p_2=1$ as the most probable value, confirming our initial assumption that $p_i=1$ for $i\geq n_\text{max}$.

    The value of $\eta$ is limited by the size of the detector (\qtyproduct{120 x 120}{\nano\meter}) as compared to the laser spot with an approximate diameter of \qty{50}{\um} used to illuminate the detector.
    In our experiment the internal detection efficiency $p_1$ is relatively low compared to what is achieved in commercial \ce{NbTiN} meandering wire SNSPDs.
    We suggest that this should be attributed to the bowtie geometry of the nanobridge detector, which is comprised of the \qtyproduct{120 x 120}{\nm} nanobridge with two tapers that angle of at \qty{45}{\degree} to the wide supply wire.
    Very close to the nanobridge, the current density is relatively high and the wire is thus still photo sensitive.
    Previously it was shown that for lower current densities, higher order events (2 or more photons) are still possible~\cite{Wang2015, Renema2012}, therefore the tapers will have an area where single photons do not cause a detection event, but higher order events will.
    In the experiment an average of the entire photo sensitive area is observed, leading to a lower effective value of $p_1$.
    We can model the total detector as a \qtyproduct{120 x 120}{\nano\meter} detector where $p_i=1$ for all photon numbers and two tapers where $p_1=0$ and $p_{i\geq2}=1$ on both side.
    Using this model, an taper area of approximately \qty{40}{\nm} long on both sides in the taper could explain the found total value of $p_1$.

    
    The nanobridge detectors studied here have a single active area comparable to the thermal healing length in \ce{NbTiN} of \qtyrange{30}{50}{\nm}~\cite{Skocpol1974}.
    Therefor multi-photon events correspond to multiple photons absorbed in the same area and it is expected that $p_2\approx1$ for $p_1>0.5$.
    Quantum detector tomography methods are agnostic and provide the parameters $\eta$ and $p_i$ that describe the detection process.
    For longer wires multi-photon processes typiccaly correspond to single photons being absorbed at different locations along the wire.
    In this scenario $p_2$ should be interpreted as a probability that a click is generated by either of the absorption events.
    This gives for the multiphoton detection probability that $p_2\geq1-\left(1-p_1\right)^2 > p_1$ for longer wires.

    In conclusion, we demonstrated that using Markov chain Monte Carlo Quantum Detector Tomography makes it is possible to separate the internal and external quantum efficiencies for devices with high internal quantum efficiencies.
    The found value for the internal quantum efficiency can be explained using the geometry of the photon detecting superconducting wire.
    Using MCMC Quantum Detector Tomography enables separating the geometric limitations on single photon detection, enabling further understanding in the working mechanisms of SNSPDs.

    \begin{acknowledgments}
    This research is made possible by financial support of the Dutch Research Council (NWO) under project 19716.
    \end{acknowledgments}

    \section*{Author Declarations}
    \subsection*{Conflict of interest}
    I.E.Z. is (part-time) employed by Single Quantum B.V. and may profit financially.
    
    \subsection*{Author Contributions}
    \textbf{F.B. Baalbergen}: Investigation (equal); Methodology (equal); Writing - original draft (lead); Writing - review \& editing (equal). 
    \textbf{I.E. Zadeh}: Investigation (equal); Methodology (equal); Writing - review \& editing (equal).
    \textbf{M.P. van Exter}: Investigation (equal); Methodology (equal); Writing - review \& editing (equal).
    \textbf{M.J.A. de Dood}: Investigation (equal); Methodology (equal); Writing - review \& editing (equal).
    
    \section*{Data Availability}
    The data that support the findings of this study are available from the corresponding author upon reasonable request.

\end{document}